\documentclass[
aps,
prl,
showpacs
amsmath,
amssymb,
floatfix,
longbibliography,
letterpaper,
lengthcheck,
superscriptaddress,
nofootinbib
]{revtex4-1}

\usepackage{graphicx}
\usepackage{rotating}
\usepackage{amsmath}
\usepackage{subfigure}
\usepackage{color}
\usepackage{soul}
\usepackage{braket}
\usepackage{hyperref}
\usepackage{cancel}
\usepackage{ulem}

\begin{document}

\title{Exact functionals for correlated electron-photon systems}

\author{Tanja Dimitrov}
\email[Electronic address:\;]{tanja.dimitrov@mpsd.mpg.de}
  \affiliation{Max Planck Institute for the Structure and Dynamics of Matter and Center for Free-Electron Laser Science Department of Physics, Luruper Chaussee 149, 22761 Hamburg, Germany}
\author{Johannes Flick}
  \email[Electronic address:\;]{johannes.flick@mpsd.mpg.de}
  \affiliation{Max Planck Institute for the Structure and Dynamics of Matter and Center for Free-Electron Laser Science Department of Physics, Luruper Chaussee 149, 22761 Hamburg, Germany}
\author{Michael Ruggenthaler}
  \email[Electronic address:\;]{michael.ruggenthaler@mpsd.mpg.de}
  \affiliation{Max Planck Institute for the Structure and Dynamics of Matter and Center for Free-Electron Laser Science Department of Physics, Luruper Chaussee 149, 22761 Hamburg, Germany}
\author{Angel Rubio}
  \email[Electronic address:\;]{angel.rubio@mpsd.mpg.de}
  \affiliation{Max Planck Institute for the Structure and Dynamics of Matter and Center for Free-Electron Laser Science Department of Physics, Luruper Chaussee 149, 22761 Hamburg, Germany}
 
\date{\today}

\begin{abstract}
For certain correlated electron-photon systems we construct the exact density-to-potential maps, which are the basic ingredients of a density-functional reformulation of coupled matter-photon problems. We do so for numerically exactly solvable models consisting of up to four fermionic sites coupled to a single photon mode. We show that the recently introduced concept of the intra-system steepening (T.Dimitrov et al., {\bf18}, 083004 NJP (2016)) can be generalized to coupled fermion-boson systems and that the intra-system steepening indicates strong exchange-correlation (xc) effects due to the coupling between electrons and photons. The reliability of the mean-field approximation to the electron-photon interaction is investigated and its failure in the strong coupling regime analyzed. We highlight how the intra-system steepening of the exact density-to-potential maps becomes apparent also in observables such as the photon number or the polarizability of the electronic subsystem. We finally show that a change in functional variables can make these observables behave more smoothly and exemplify that the density-to-potential maps can give us physical insights into the behavior of coupled electron-photon systems by identifying a very large polarizability due to ultra-strong electron-photon coupling.
\end{abstract}

\date{\today}

\maketitle

\section{Introduction}
Recent experiments~\cite{slootsky2014,orgiu2015, shalabney2015,shalabney2015a,george2015,george2016,thomas2016,chikkaraddy2016,ebbesen2016} at the interface of quantum chemistry, material science and quantum optics allow to tailor the physical and chemical properties of the system by coupling light strongly to the matter, e.g. by placing it in an optical cavity. The theoretical description of such experiments requires a full quantum treatment of the entire system including the electronic matter and the electromagnetic field. Common electronic-structure methods, such as density-functional theory (DFT)~\cite{kohn1965,gross1984} allow to efficiently describe the quantum nature of the electrons while the electromagnetic field is {treated as a static and fixed external perturbation.} To also include the electromagnetic field explicitly and thus being able to describe, e.g. chemical systems in an optical cavity, time-dependent and ground-state DFT have been recently generalized to correlated electron-photon system ~\cite{Ruggenthaler2011,Tokatly2013,Ruggenthaler2014,Ruggenthaler2015}. This new density-functional framework for coupled matter-photon problems has been termed quantum electrodynamical density-functional theory (QEDFT)~\cite{Ruggenthaler2014, flick2015, flick2016a}. Similar to DFT, QEDFT is an exact framework to describe the many-body problem~\cite{Ruggenthaler2014,Ruggenthaler2015}. Both frameworks exploit the one-to-one correspondence between the internal and external variables that are formally connected via a Legendre transformation. As a consequence of these so-called \textit{density maps}, one can determine every observable of the quantum system as a functional of the internal variables only. While in DFT the internal variable is the one-particle electron density (conjugate to the external scalar potential), in QEDFT we have two internal variables (one for the electrons and one for the photons). {These variables} depend on the form of the electron-photon Hamiltonian under considerations \cite{Ruggenthaler2014}. 
In DFT, to calculate the physical density of a many-body system and thus avoid the numerically infeasible correlated many-body wave function, one usually employs the Kohn-Sham scheme~\cite{kohn1965}. In this approach the $N$-particle Schr\"odinger equation is replaced by $N$ coupled, non-linear one-particle equations, which are numerically tractable. The price to pay is that these effective particles are subject to an in general unknown xc potential, which makes up for all the missing many-body effects. Also in QEDFT we can replace the full electron-photon Schr\"odinger equation by coupled, non-linear one-particle equations. The electronic subsystem is again described by equations for single particles that are subject to a xc field. In this case, however, the effective field does not only contain contributions from many-body effects due to the electron-electron interaction but also from many-body effects due to the photon-electron interactions \cite{Tokatly2013,Ruggenthaler2014}. Further, the photonic subsystem is described by an inhomogeneous Maxwell equation, where the inhomogeneity is usually given explicitly by the electronic subsystem \cite{Tokatly2013,Ruggenthaler2014}. 
\\
In practice, calculations within the new QEDFT framework require reliable approximations to the unknown xc potentials. Herein, QEDFT profits from the long-standing search~\cite{Perdew2001} in DFT for more reliable xc potentials that efficiently mimic the electron-electron interaction. While common xc functionals can be used to describe the many-body effects due to electron-electron interactions, new functionals that mimic the electron-photon interaction have to be developed. {In this work, we are concerned about the xc potential of the light-matter interaction, i.e. the potential an electron encounters due to its coupling to the electromagnetic field.} For the electron-photon contributions first approximations for the xc potential along the lines of the optimized effective potential (OEP) approximation have been already demonstrated to be practical~\cite{Pellegrini2015,flick2016a}.
If, however, common approximations for the electron-electron many-body effects are used, then clearly QEDFT will face the same challenges as standard DFT when systems with strong electron-electron correlations are considered. To better understand such situations in DFT, the impact of static correlation and localization for different exact density maps has been analyzed in a recent work~\cite{Dimitrov2016}. By investigating specific integrated quantities of these maps, e.g. the density difference between two parts of the system $\delta n$, it has been shown that static correlation and localization can be quantified by the concept of \textit{intra-system steepening}. For $\delta n$ a step can be found that becomes steeper with increasing correlation in the system. This feature translates to different functionals of the density, and corresponds to the full real-space behavior of steps and peaks in the exact xc potential~\cite{Buijse1989,Gritsenko1994,Gritsenko1996,fuks2016}. In QEDFT we have besides the electron-electron correlations also electron-photon correlations. And also for them according step and peak structures in the xc potential appear in real space and pose a challenge for constructing approximate xc potentials that are reliable for strong electron-photon correlations \cite{flick2015}. Consequently, can we analyze the correlation and localization in a similar manner for coupled electron-photon system, and is the intra-system steepening a general feature of correlated systems?
\\
In this work, we construct the exact density-to-potential maps of ground-state QEDFT~\cite{Ruggenthaler2015} and {examine} the intra-system steepening related to the real-space properties of the exact xc potentials {for correlated electron-photon systems}. For electron-photon model systems we show that the localization of the electrons and the displacement of the photon mode depends on the ratio between the kinetic energy and the coupling term between electrons and photons. Features of this intra-system steepening can also be found in other observables, such as the photon number. A change in functional variables though, e.g., by going from the external to the conjugate internal variables, can make the behavior of these observables more regular. We further show how the validity of the mean-field approximation to the electron-photon coupling can be investigated by analyzing the intra-system steepening. Finally we highlight how density-potential maps in electron-photon systems can be used also outside of QEDFT to analyze the properties of physical systems by investigating the polarizability of an electron-photon system when increasing the coupling strength.
\section{Exact maps and the Kohn-Sham construction in QEDFT}
QEDFT allows to describe the quantum nature of electrons and photons on the same footing by reformulating coupled matter-photon problems in an exact quantum fluid description. In the following we consider the interaction of a system of $n_e$ electrons, e.g., a molecule in Born-Oppenheimer approximation~\cite{flick2016b}, with $n_p$ quantized modes of a photon field. A typical experimental situation would be to place the matter system inside an optical cavity, where only specific frequencies are assumed to interact with the multi-particle system. Such a situation can be described by employing the following Hamiltonian~\cite{Tokatly2013, Pellegrini2015, flick2016a}
\begin{align}
\label{eq:qed-chemistry-hamiltonian}
\hat{H}(t) &= \hat{H}_{e}(t) + \hat{H}_p(t)\\ 
\hat{H}_e(t)&=\sum_{i=1}^{n_e}\left(-\frac{\hbar^2}{2m}\vec{\nabla}_i^2+v_\text{ext}(\textbf{r}_i,t)\right)\nonumber\\
&+\frac{e^2}{4 \pi\epsilon_0}\sum_{ij,i>j}\frac{1}{\left|\textbf{r}_i-\textbf{r}_j\right|}\label{eq:ham_ee}\\
\hat{H}_p(t) &= \frac{1}{2}\sum\limits_{\alpha=1}^{n_p}\left[\hat{p}^2_\alpha+\omega_\alpha^2\left(\hat{q}_\alpha - \frac{\boldsymbol \lambda_\alpha} {\omega_\alpha} \cdot{e\textbf{R}} \right)^2\right]  +\frac{j^{(\alpha)}_\text{ext}(t)}{\omega_\alpha}\hat{q}_\alpha.\label{eq:ham-photons}\\
\label{eq:dipole_operator}
{\textbf{R}} & = \sum_{i=1}^{N_e} \textbf{x}_i,
\end{align}
where ${\textbf{R}}$ refers to the electronic dipole operator. {Note, in this work, we neglect electron-nuclear interactions by working in the clamped-ion approximation. Therefore, the Hamiltonian given above only couples the electromagnetic field to the electrons. However, extending the work to the interaction between the ions and the field is straightforward, but would make the discussion in the present work more cumbersome.} Besides the usual Schr\"odinger Hamiltonian $\hat{H}_{e}(t)$ that describes the charged-particle system, we now also have $n_p$ photon modes with frequencies $\omega_{\alpha}$ that are coupled in dipole approximation with the electronic system. 
Here the photon momenta {$\hat{p}_\alpha=\frac{1}{i}\sqrt{\frac{\omega_\alpha}{2}}\left(\hat{a}_\alpha-\hat{a}_\alpha^\dagger\right)$} in terms of the usual creation and annihilation operators are connected to the magnetic field for mode $\alpha$, and $\hat{q}_{\alpha} =\sqrt{\frac{1}{2\omega_{\alpha}}}\left(\hat{a}_{\alpha}+\hat{a}_{\alpha}^\dagger\right)$ is proportional to the electric displacement field. Therefore we have to subtract the polarization of the electronic system such that $(\omega_{\alpha} \hat{q}_{\alpha} - {\boldsymbol \lambda_\alpha} \cdot{e\textbf{R}})$ corresponds to the electric field. The coupling strength is $|{\boldsymbol \lambda_\alpha}|$ and ${\boldsymbol \lambda_\alpha}/|{\boldsymbol \lambda_\alpha}|$ is the polarization vector. Further, $j^\alpha_\text{ext}(t)$ corresponds to an external dipole moment that drives mode $\alpha$.
\\
To reformulate the above problem we employ a bijective mapping between the external variables of the system, i.e., $v_\text{ext}(\textbf{r},t)$ and $j^{(\alpha)}_\text{ext}(t)$, and the conjugate internal variables~\cite{Tokatly2013, Ruggenthaler2011, Ruggenthaler2014, Ruggenthaler2015} given here by $n(\textbf{r},t)$ and $q_{\alpha}(t)$, i.e.,
\begin{align}
(v_\text{ext}(\textbf{r},t),j^{(\alpha)}_\text{ext}(t) ) \underset{1:1}{\longleftrightarrow} \left(n(\textbf{r},t),q_\alpha(t) \right).
\end{align}
While in principle this mapping allows to calculate the exact internal variables by solving a local-force equation for the charge density non-linearly coupled to a classical Maxwell equation~\cite{Tokatly2013, Ruggenthaler2011, Ruggenthaler2014, Ruggenthaler2015}, in general we do not know the exact form of the momentum-stress and interaction forces in such equations~\cite{tokatly2005a,tokatly2005b}. So in practice we have to use approximations. The standard way to devise such approximations is the use of a non-interacting auxiliary system, a so-called Kohn-Sham system~\cite{ruggenthaler2015b}. In the Kohn-Sham scheme the difference in forces between the non-interacting and interacting system is subsumed in a mean-field term and the unknown xc potential. In the case of coupled electron-photon systems the mean-field contribution is the classical Maxwell field, which has the usual longitudinal Hartree contribution and now also transversal terms, and the xc potential contains the electron-electron and electron-photon many-body effects. Neglecting the electron-photon many-body effects in the xc potential in the case of coupled electron-photon systems leads to the mean-field potential that is identical to a classical Maxwell-Schr\"odinger simulation~\cite{yabana2012,li2016}. 
\\
Approximations to the xc potential of the coupled electron-photon system face similar problems to the ones of purely electronic systems. When increasing the correlation, i.e. increasing the coupling strength $|{\boldsymbol \lambda_\alpha}|$, the accuracy of the mean-field or the exchange-only OEP~\cite{Pellegrini2015} decreases. To improve and construct approximations that can treat strong-coupling situations more accurately we need a better understanding of the electron-photon contributions in the strong-coupling limit. To this end we  explicitly construct and investigate the exact fundamental maps that underly the framework of ground-state QEDFT. As model system, we choose the Rabi-Hubbard model, i.e. a few-site model coupled to a single photon mode. We consider three different setups (i) a single electron on two sites, where the electron-electron interaction favoring the localization in the system is equal to zero. (ii) Two electrons on two-sites, where we model the electron-electron repulsion by a Hubbard interaction term. We analyze both maps in the resonant limit for different coupling strength. (iii) Four electrons on four sites, here we connect the intra-system steepening and the modification of the electric polarizability for such systems. 
\section{Two-site Rabi-Hubbard Model}
\begin{figure}[!htbp]
\centerline{\includegraphics[width=0.5\textwidth]{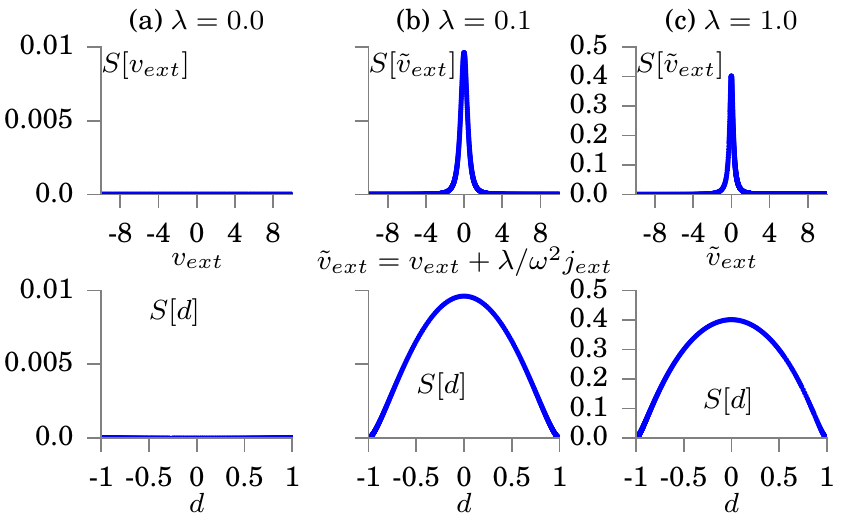}}
\caption{Correlation entropy as function of the polaritonic external variables $\tilde{v}_\text{ext}$ in the first row and as function of the electron density $d$ in the second row.}
\label{fig:van_neumman}
\end{figure}

\begin{figure}[!htbp]
\centerline{\includegraphics[width=0.5\textwidth]{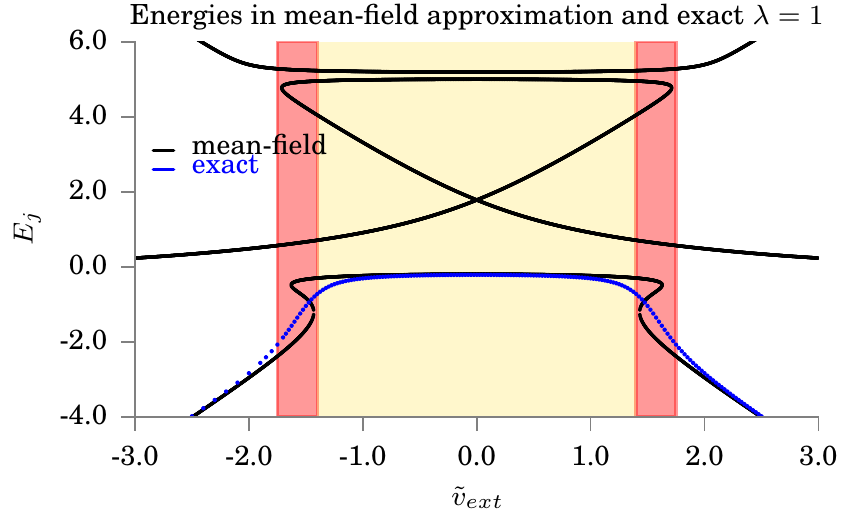}}
\caption{Eigenvalues of the exact Hamiltonian of Eq.~\ref{eq:ham_ee} versus the mean field approximation of the electronic energy of  Eq.~\ref{eq:ham_e_mf} for the two-site Rabi-Hubbard model. Indicated in black are all mean-field eigenvalues in the shown range, while in blue the exact ground-state value is shown.} 
\label{fig:mf-eig}
\end{figure}

\begin{figure}[!htbp]
\centerline{\includegraphics[width=0.5\textwidth]{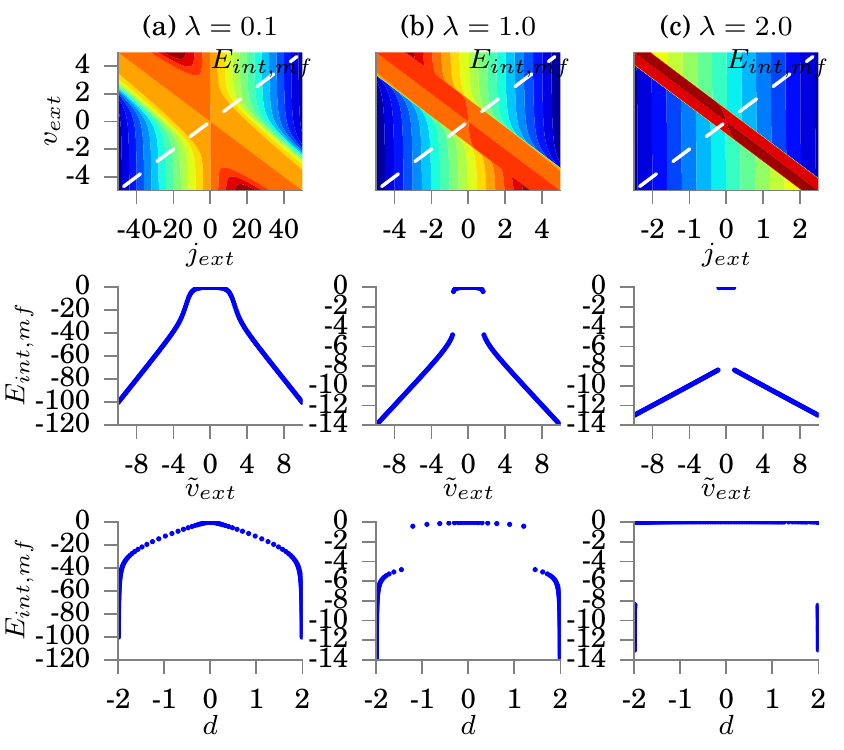}}
\caption{Two electrons with $U_0=5$ on a two-site lattice: the interaction energy $E_{int,mf} = \omega {q} {d}$ in mean-field approximation as function of the external variables $(v_\text{ext},j_\text{ext})$ is shown in the first row. The second row shows the antidiagonal cut of $E_{int,mf}(v_\text{ext},j_\text{ext})$ as indicated by the dashed line in the upper plot. The third row shows the diagonal cut of $E_{int,mf}(v_\text{ext},j_\text{ext})$. All plots are shown for three different coupling strengths (a) $\lambda=0.1$, (b) $\lambda=1$, and (c) $\lambda=2$.} 
\label{fig:e-int-v-j-mf}
\end{figure}

\begin{figure}[!htbp]%
\centerline{\includegraphics[width=0.5\textwidth]{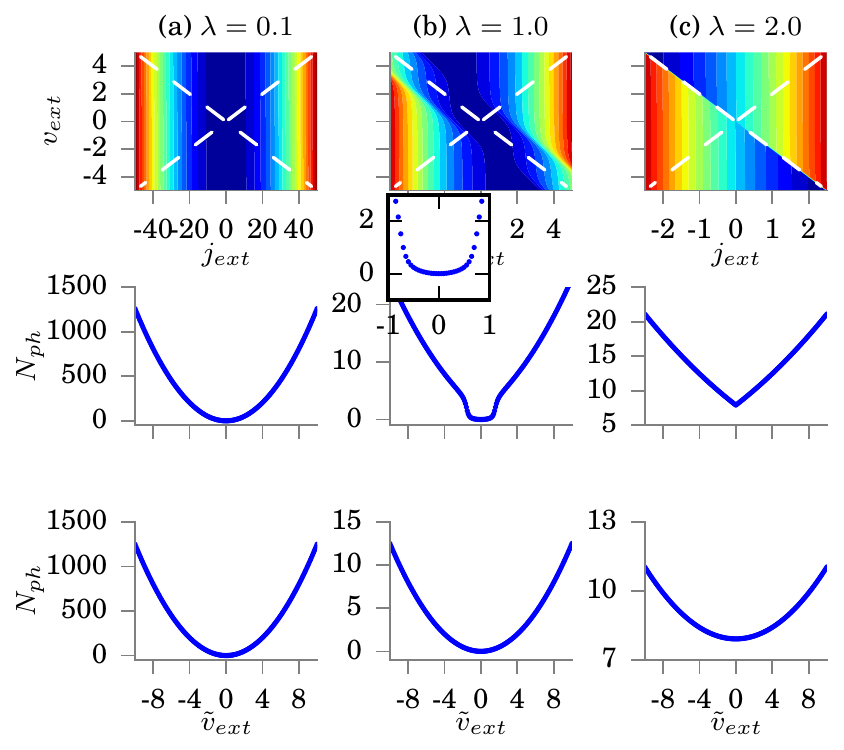}}%
\caption{Two electrons with $U_0=5$ on a two-site lattice: the photon number $N=\langle\hat{a}^\dagger\hat{a} \rangle$ as function of the external variables $(v_\text{ext},j_\text{ext})$ is shown in the first row. The second row shows the antidiagonal cut of $N(v_\text{ext},j_\text{ext})$ as indicated by the dashed line in the upper plot. The third row shows the diagonal cut of $N(v_\text{ext},j_\text{ext})$. All plots are shown for three different coupling strengths (a) $\lambda=0.1$, (b) $\lambda=1$, and (c) $\lambda=2$.}%
\label{fig:N-ph}%
\end{figure}

The Rabi model~\cite{Pellegrini2015, Shore1993}, which consists of one electron on two sites coupled to one photonic mode, has been heavily investigated in the context of light-matter interactions~\cite{Braak2011}, e.g. recently in the context of photon blockade~\cite{boite2016}. In this work, we employ a generalized Rabi model with $n_s$ sites and that can host up to $2n_s$ interacting electrons (Rabi-Hubbard model). The corresponding model Hamiltonian reads as follows\footnote{We note here, that in the continuum limit, the dipole self-interaction term $(\boldsymbol \lambda_\alpha \cdot{e\textbf{R}})^2/2$ term becomes important, see e.g. the discussion in Ref.~\cite{flick2016a}. However, in the two-site case the dipole self-energy corresponds to a constant energy shift that we neglect in the discussion of the two-site model.}
\begin{align}
\hat{H}_0&=-t_0 \sum\limits_{i=1,\sigma=\uparrow,\downarrow}^{n_s-1}\left(\hat{c}_{i,\sigma}^\dagger\hat{c}_{i+1,\sigma} + \hat{c}_{i+1,\sigma}^\dagger\hat{c}_{i,\sigma}\right)\nonumber\\ &+U_0\sum\limits_{i=1}^{n_s}\hat{n}_{i,\uparrow}\hat{n}_{i,\downarrow}
+ \omega \hat{a}^{\dagger}\hat{a} - {\omega}\lambda \hat{q} \hat{d} +\frac{j_\text{ext}}{\omega}\hat{q} \nonumber\\
&+ (\lambda\hat{d})^2/2+v_\text{ext}\hat{d}
\label{eq:ham_e_ph}
\end{align}
where the photon displacement operator is given by $\hat{q} =\sqrt{\frac{1}{2\omega}}\left(\hat{a}+\hat{a}^\dagger\right)$ (the photon momentum operator $\hat{p}=\frac{1}{i}\sqrt{\frac{\omega}{2}}\left(\hat{a}-\hat{a}^\dagger\right)$) and $\lambda$ introduces a coupling between the electronic and photonic part of the system. The electronic part is described by the standard Hubbard model with the on-site parameter $U_0$, the hopping matrix element $t_0$, and the operators $\hat{c}^\dagger_{i,\sigma}$ and $\hat{c}_{i,\sigma}$ that create or destroy an electron with spin $\sigma$ on site $i$. The electron density operator on site $i$ is given by $\hat{n}_i=\sum_\sigma \hat{c}^\dagger_{i,\sigma}\hat{c}_{i,\sigma}$. We furthermore specify the dipole moment of the electronic system by $d = \int \hat{d}n(r)dr$, for two sites this corresponds to $\delta n=n_1-n_2$, i.e. the density difference between both sites in the lattice and $d=\delta n$
\footnote{We emphasize that the {two-site} Rabi-Hubbard Hamiltonian as in Eq.~\ref{eq:ham_e_ph} is exactly identical to a Holstein-Hubbard Hamiltonian that is routinely used in the electron-phonon community, e.g. discussed in Refs.~\cite{sakkinen2014, sakkinen2015,Berciu2007}.}.\\
In the case of the above Hamiltonian of Eq.~\ref{eq:ham_e_ph} the pair of conjugate variables are $(v_\text{ext},j_\text{ext})$ and $(d=\langle \hat{d}\rangle,q=\langle \hat{q}\rangle)$ \cite{Ruggenthaler2015}. A simple way to see that this is true from a purely electronic DFT perspective and that helps to interpret the external term $j_\text{ext}$ is by performing a unitary transformation of the above Hamiltonian. With the coherent-shift operator $U[j_\text{ext}] = \exp(i j_\text{ext}\hat{p}/\omega^3)$ we can recast the Hamiltonian of Eq.~\ref{eq:ham_e_ph} into the unitarily equivalent form
\begin{align}
\label{eq:ham0_trans}
\hat{H}_0'&=\hat{U}^{\dagger}\hat{H}_{0}\hat{U}\\
&=-t_0 \sum\limits_{i,\sigma=\uparrow,\downarrow}^{n_s-1}\left(\hat{c}_{i,\sigma}^\dagger\hat{c}_{i+1,\sigma} + \hat{c}_{i+1,\sigma}^\dagger\hat{c}_{i=1,\sigma}\right) +U_0\sum\limits_{i=1}^{n_s}\hat{n}_{i,\uparrow}\hat{n}_{i,\downarrow}\nonumber\\
&+ \omega \hat{a}^{\dagger}\hat{a} - {\omega}\lambda \hat{q} \hat{d} +(\lambda\hat{d})^2/2+ (v_\text{ext} + \frac{\lambda}{\omega^2} j_\text{ext})\hat{d} - \frac{1}{2 \omega^4} j_\text{ext}^2.\nonumber
\end{align}
Thus, we see that the external dipole $j_\text{ext}$ can be recast into an external potential on the electrons by a unitary transformation. 
\begin{figure}[htbp]
\centerline{\includegraphics[width=0.5\textwidth]{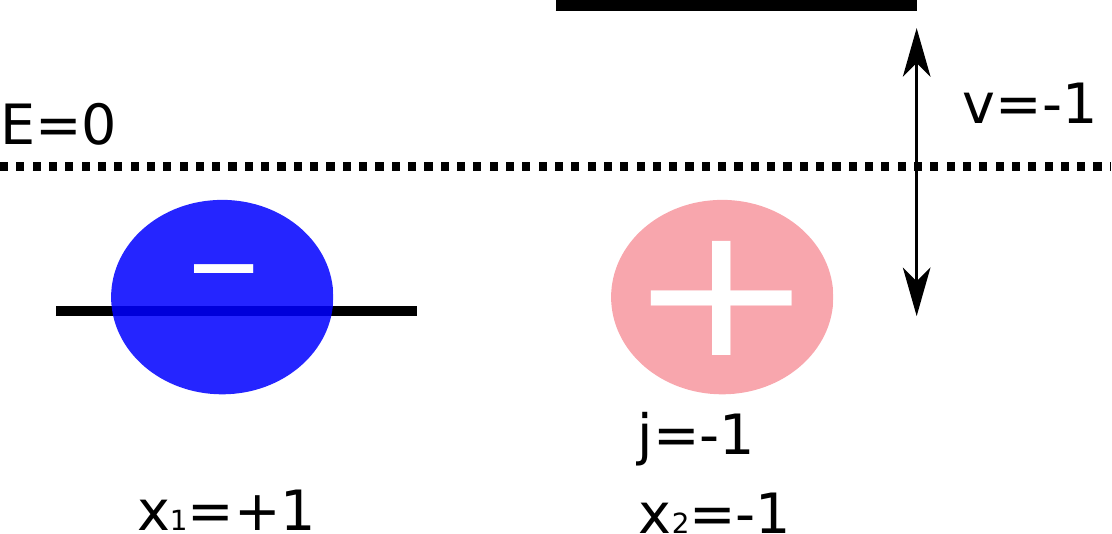}}
\caption{Schematic view on the two-side model: A negative external potential $v_\text{ext}=-1$ introduces an energy difference between the two-sites. The electrons in the electronic ground-state become localized on the left side. The external variable for the photon field, $j_\text{ext}$ can be interpret as a classical charge that generates an external potential as well. If $j_\text{ext}= -\frac{\omega^2}{\lambda}$ then the electron is again delocalized.}
\label{fig:schematic-model}
\end{figure}
Take, for instance, the case of the two-site problem  Rabi-Hubbard model as depicted in Fig.~\ref{fig:schematic-model}. If $j_\text{ext}=0$ and a negative external potential $v_\text{ext} <0$ acts on the system, the external potential localizes the electron on one site. The external dipole for the photons $j_\text{ext}$ introduces a classical positive charge to the system that can counterbalance the effect of the external potential $v_\text{ext}$. With the usual Hohenberg-Kohn theorem we know that for any external potential $\tilde{v}_\text{ext}=(v_\text{ext} + \frac{\lambda}{\omega^2} j_\text{ext})$ there is one and only one ground-state wave function $\Psi_0'$ associated. And from this ground-state we find the corresponding unique wave function of the original problem by $\Psi_0 = D[-j_\text{ext}]\Psi_0'$. Thus purely electronic properties can be reconstructed from the situation with $j_\text{ext}=0$, while the photonic observables will in general depend in a non-trivial manner on the $j_\text{ext}$. Further, as can be deduced from the equations of motions for the photonic systems (e.g. Eq.~2 in Ref.~\cite{flick2015}.), we can establish a direct connection between $q$ and $d$ and $j_\text{ext}$ for the ground-state ($\frac{\partial}{\partial t}q=\frac{\partial^2}{\partial t^2}q=0$)
\begin{align}
q = \frac{\lambda}{\omega}d - \frac{1}{\omega^3}j_\text{ext}.
\label{eq:q-n-j}
\end{align}
Using the external variables $v_\text{ext}$ and $j_\text{ext}$, we stepwise screen the  external potential of the photons and electrons. 
For each fixed pair of the external potential $(v_\text{ext},j_\text{ext})$, we diagonalize the Hamiltonian using exact diagonalization~\cite{flick2014, Dimitrov2016} given in Eq.~\ref{eq:ham_e_ph} and obtain the corresponding ground-state wave function of the system, in the following denoted by $\Psi_0(v_\text{ext},j_\text{ext})$. Using the exact wave function, we have access to the conjugated set of variables, i.e. $(d,q)$, by evaluating the corresponding expectation values 
\begin{align}
d=\bra{\Psi_0^{(v_\text{ext},j_\text{ext})}}{\hat{d}}\ket{\Psi_0^{(v_\text{ext},j_\text{ext})}}
\end{align}
and 
\begin{align}
q=\bra{\Psi_0^{(v_\text{ext},j_\text{ext})}}\hat{q}\ket{\Psi_0^{(v_\text{ext},j_\text{ext})}}
\end{align}
corresponding to the electronic dipole and the photonic displacement coordinate. Screening the parameters $v_\text{ext}$ and $j_\text{ext}$ allows us to construct the complete map between the conjugated set of variables.\\
For general many-body calculations, we can use the Kohn-Sham approach~\cite{kohn1965} to simulate the interacting many-body problem by solving equations for non-interacting particles. In the electron-photon situation that is presented here, we encounter two interaction terms, i.e. the electron-electron interaction modeled by a Hubbard on-site interaction and the electron-photon interaction. In general, we can setup a Kohn-Sham system for non-interacting electrons as presented in Refs.~\cite{Ruggenthaler2014,flick2015}. However, in this paper we focus on the effects of the electron-photon interaction on the density-to-potential maps and we therefore include the electron-electron interaction in the Kohn-Sham system explicitly. Thus, the Kohn-Sham system reads in the case of a two-site lattice as follows
\begin{align}
   \hat{H}_{fm,KS}&=-t_0 \sum\limits_{\sigma=\uparrow,\downarrow}\left(\hat{c}_{1,\sigma}^\dagger\hat{c}_{2,\sigma} + \hat{c}_{2,\sigma}^\dagger\hat{c}_{1,\sigma}\right)\nonumber\\
   &+U_0\sum\limits_{i=1,2}\hat{n}_{i,\uparrow}\hat{n}_{i,\downarrow}+ v_\text{S}\hat{d}
   \label{eq:ham_e_ks}\\
      \hat{H}_{ph,KS}&= \omega \hat{a}^{\dagger}\hat{a} +\frac{j_\text{S}}{\omega}\hat{q}.
   \label{eq:ham_ph_ks}
\end{align}
The hereby emerging effective Kohn-Sham potential $v_\text{S}$ and the effective current $j_\text{S}$ are chosen such that the ground-state density is equal in the Kohn-Sham systems of Eq.~\ref{eq:ham_e_ks}-\ref{eq:ham_ph_ks} and the full interacting problem of Eq.~\ref{eq:ham_e_ph}. While the effective current $j_\text{S}$ is known explicitly~\cite{Ruggenthaler2014,flick2014}, i.e. $j_\text{S}=-{\omega}^2\lambda \hat{q} {d} +{j_\text{ext}}$, the effective potential $v_\text{S}$ has to be approximated. To this end, we divide $v_\text{S}$ as follows
\begin{align}
v_\text{S}=v_\text{ext} + v_\text{M} + v_\text{xc},
\label{eq:exchange-correlation}
\end{align}
{where $v_\text{M}$ and $v_\text{xc}$ describe the mean-field part and the xc part, respectively.}\\
The simplest approximation to the fully coupled problem and the starting point for the Kohn-Sham construction in the electron-photon case is the mean-field approximation~\cite{flick2015} {that is given by $v_M=- {\omega}\lambda {q} \hat{d}$} and leads to the following Hamiltonian in the case of a two-site lattice
\begin{align}
   \hat{H}_{fm,0}&=-t_0 \sum\limits_{\sigma=\uparrow,\downarrow}\left(\hat{c}_{1,\sigma}^\dagger\hat{c}_{2,\sigma} + \hat{c}_{2,\sigma}^\dagger\hat{c}_{1,\sigma}\right) +U_0\sum\limits_{i=1,2}\hat{n}_{i,\uparrow}\hat{n}_{i,\downarrow}\nonumber\\
   &- {\omega}\lambda {q} \hat{d}  + v_\text{ext}\hat{d}
   \label{eq:ham_e_mf}\\
      \hat{H}_{ph,0}&= \omega \hat{a}^{\dagger}\hat{a} - {\omega}\lambda \hat{q} {d} +\frac{j_\text{ext}}{\omega}\hat{q},
   \label{eq:ham_ph_mf}
\end{align}
where $d=\langle d\rangle$ and $q=\langle q\rangle$. To obtain the mean-field ground state, Eqns.~\ref{eq:ham_e_mf}-\ref{eq:ham_ph_mf} have to be solved either self-consistently, or Eq.~\ref{eq:q-n-j} can be exploited leading to the following electronic equation
\begin{align}
   \hat{H}_{fm,0}&=-t_0 \sum\limits_{\sigma=\uparrow,\downarrow}\left(\hat{c}_{1,\sigma}^\dagger\hat{c}_{2,\sigma} + \hat{c}_{2,\sigma}^\dagger\hat{c}_{1,\sigma}\right) +U_0\sum\limits_{i=1,2}\hat{n}_{i,\uparrow}\hat{n}_{i,\downarrow}\nonumber\\
   &- \lambda^2 d \ \hat{d}  + \frac{\lambda}{\omega^2} j_\text{ext} \hat{d} + v_\text{ext}\hat{d}
   \label{eq:ham_e_mf2}.
\end{align}
In these equations, we apply the classical approximation only to the electron-photon interaction, while the electron-electron interaction is treated fully correlated. We may expect that such a approximation works well for the studied model in the weak-coupling regime and in the limit of infinite coupling~\cite{Pellegrini2015}.\\
To construct the exact $v_{xc}$ of Eq.~\ref{eq:exchange-correlation} beyond the mean-field approximation, we can, for instance, use the Heisenberg equation of motion to find the connection between the electronic density $d$ and $v_\text{S}$ for the Kohn-Sham system and between $d$ and $v_\text{ext}$ in the many-body problem. These equation read for the ground state as follows
\begin{align}
d[v_\text{ext},j_\text{ext}]&=\langle\frac{\omega\lambda \hat{q}-v_\text{ext}}{t_0}\sum\limits_{\sigma=\uparrow,\downarrow}\left(\hat{c}_{1,\sigma}^\dagger\hat{c}_{2,\sigma} + \hat{c}_{2,\sigma}^\dagger\hat{c}_{1,\sigma}\right)\rangle\nonumber\\
&-\frac{1}{2t_0}U_0\langle\left(\hat{c}_{1,\uparrow}^\dagger\hat{c}_{2,\uparrow} + \hat{c}_{2,\uparrow}^\dagger\hat{c}_{1,\uparrow}\right)\left(\hat{n}_{1,\downarrow}-\hat{n}_{2,\downarrow}\right)\rangle\nonumber\\
&-\frac{1}{2t_0}U_0\langle\left(\hat{c}_{1,\downarrow}^\dagger\hat{c}_{2,\downarrow} + \hat{c}_{2,\downarrow}^\dagger\hat{c}_{1,\downarrow}\right)\left(\hat{n}_{1,\downarrow}-\hat{n}_{2,\downarrow}\right)\rangle\\
d[v_\text{S},j_\text{S}]&=-\frac{v_\text{S}}{t_0}\sum\limits_{\sigma=\uparrow,\downarrow}\langle\left(\hat{c}_{1,\sigma}^\dagger\hat{c}_{2,\sigma} + \hat{c}_{2,\sigma}^\dagger\hat{c}_{1,\sigma}\right)\rangle\nonumber\\
&-\frac{1}{2t_0}U_0\langle\left(\hat{c}_{1,\uparrow}^\dagger\hat{c}_{2,\uparrow} + \hat{c}_{2,\uparrow}^\dagger\hat{c}_{1,\uparrow}\right)\left(\hat{n}_{1,\downarrow}-\hat{n}_{2,\downarrow}\right)\rangle\nonumber\\
&-\frac{1}{2t_0}U_0\langle\left(\hat{c}_{1,\downarrow}^\dagger\hat{c}_{2,\downarrow} + \hat{c}_{2,\downarrow}^\dagger\hat{c}_{1,\downarrow}\right)\left(\hat{n}_{1,\downarrow}-\hat{n}_{2,\downarrow}\right)\rangle,
\end{align}
where in the many-body problem, the many-body wave function has to be employed to calculate observables, while in the Kohn-Sham system the factorizable Kohn-Sham wave function is employed. Since the electronic density $d$ is by construction equal in the interacting system and the exact Kohn-Sham system, if the exact Kohn-Sham potential $v_\text{S}$ is used, we find for the density-to-potential maps $d[v_S]=d[v_\text{ext}]$. By using the inverse mapping, i.e. $v_\text{ext}[d,q]$, we can construct the exact xc potential of Eq.~\ref{eq:exchange-correlation} using~\cite{ruggenthaler2015b}
\begin{align}
v_\text{xc}^{\lambda}[d,q]=v_\text{ext}^{\lambda=0}[d,q]-v_\text{ext}^{\lambda}[d,q]-v_\text{M}^{\lambda}[d,q].
\end{align}
In the following, we construct the exact density-to-potential maps of $d[v_\text{ext},j_\text{ext}]$ and $v_\text{xc}^\lambda[n,q]$ to get insights how the electron-photon interaction influences the electronic system and draw conclusions on approximations for corresponding xc potential.\\
We start discussing the Rabi-Hubbard model in setup (i), where a single electron is coupled to the photon mode of frequency $\omega=1$.
\begin{figure}[!htbp]
\centerline{\includegraphics[width=0.5\textwidth]{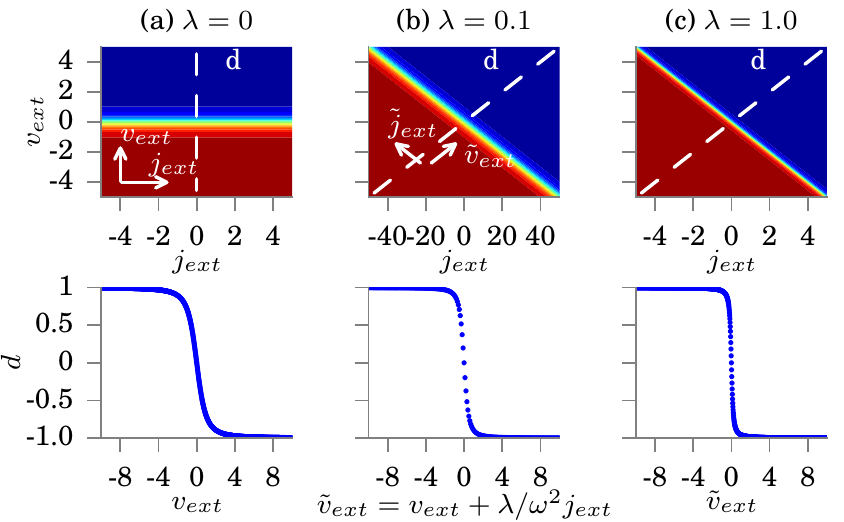}}
\caption{Single electron on two a two-site lattice: the electron density ${d}$ as function of the external variables $(v_\text{ext},j_\text{ext})$ is shown in the first row. The second row shows the cut of ${d}(v_\text{ext},j_\text{ext})$ as indicated by the dashed line in the upper plot both for different coupling strength of (a) $\lambda=0$, (b) $\lambda=0.1$, and (c) $\lambda=1$.} 
\label{fig:dn-p1}
\end{figure}
The first situation we analyze is, when the electron and the photons do not couple, see Fig.~\ref{fig:dn-p1} (a) ($\lambda=0$). In this case varying $j_\text{ext}$ has no effect on the density-to-potential map. Therefore, the density-to-potential map $d[v_\text{ext}]$ is determined by the external potential $v_\text{ext}$ alone. The dependency of $d[v_\text{ext}]$ on $v_\text{ext}$ is shown in the lower plot. We find a continuous and rather smooth mapping. Since, we have restricted ourselves to a single electron, the dipole corresponding to the density difference between both sites $d$ can have values in between $[-1,1]$. In Fig.~\ref{fig:dn-p1} (b), we now introduce a finite $\lambda$, here $\lambda=0.1$.
\begin{figure}[!htbp]
\centerline{\includegraphics[width=0.5\textwidth]{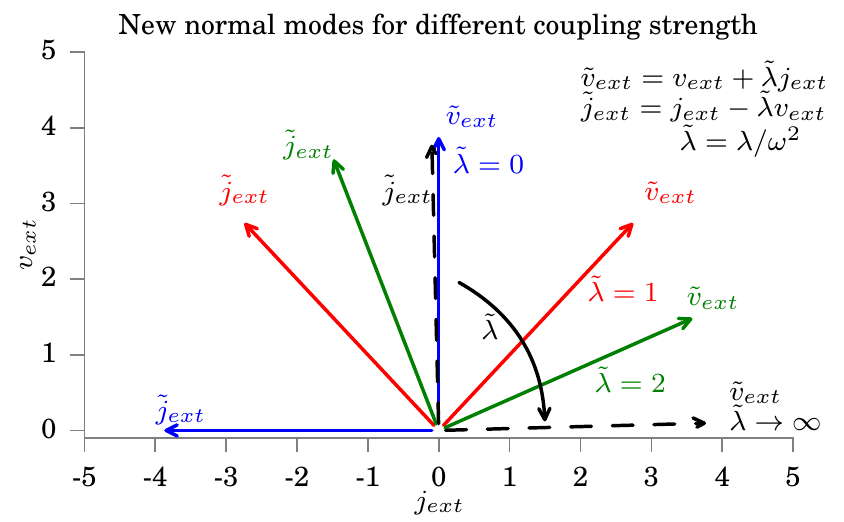}}
\caption{New normal modes appearing. Increasing electron-photon coupling strength leads to the rotation of coordinate system. }
\label{fig:normal_modes}
\end{figure}
In Fig.~\ref{fig:dn-p1} (b), we plot the two-dimensional density-to-potential map $d[v_\text{ext},j_\text{ext}]$ for $v_\text{ext}=[-5,5]$ and $j_\text{ext}=[-50,50]$. The first emerging feature in the plot is that two new normal modes appear~\cite{flick2016a,flick2016b}, i.e. the photon and electron degrees of freedom become correlated. This electron-photon correlation tildes the map as shown in Fig.~\ref{fig:normal_modes}. The rotation can be  constructed by $\tilde{v}_\text{ext}=v_\text{ext}+\lambda/\omega^2j_\text{ext}$ and corresponds to the transformation using the coherent-shift operator as in Eq.~\ref{eq:ham0_trans}. The diagonal cut in the plot is the new polaritonic degree of freedom that is shown in the plot on the bottom. We find a broad smearing of the density-to-potential map. Fig.~\ref{fig:dn-p1} (c) shows the map for $\lambda=1$. The plot is shown for $v_\text{ext}=[-5,5]$ and $j_\text{ext}=[-5,5]$, hence the photon external variable is narrower. In comparison to $\lambda=0.1$, we find a steepening of the gradient in the density-to-potential plot that we have earlier introduced as intra-system steepening~\cite{Dimitrov2016}. To highlight the connection of the steepening to electronic correlation, Fig.~\ref{fig:van_neumman} shows the correlation entropy for the one-electron system, i.e. a good measure for the static correlation and indicates how well the ground-state wave function is approximated by a single Slater determinant. The correlation entropy is given by
\begin{align}
\label{eq:vN}
S=\sum_{j=1}^{\infty}n_j \text{ln} n_j \text{,}
\end{align}
where the occupation numbers $n_j$ are the eigenvalues of the reduced one-body density matrix~\cite{Loewdin1955} that is given in terms of the many-body wave function $\Psi(\vec{x},\vec{x}_2,...,\vec{x}_N)$ as
\begin{align}
\rho_{1\text{RDM}}(\vec{x},\vec{x}')&=\\\nonumber\int d^3\vec{x}_2...d^3\vec{x}_N &\Psi^{\ast}(\vec{x},\vec{x}_2,...,\vec{x}_N)\Psi(\vec{x}',\vec{x}_2,...,\vec{x}_N)\text{.}
\end{align}
In spectral representation, the reduced density matrix can be written in terms of its eigenfunctions and eigenvalues as~\cite{Dimitrov2016}
\begin{align}
\rho_{1\text{RDM}}(\vec{x},\vec{x}')=\sum\limits_{j}n_j\phi^{\ast}_j(\vec{x})\phi_j(\vec{x}')\text{.}
\end{align}
In Fig.~\ref{fig:van_neumman} the correlation entropy increases with the coupling between the photonic and electronic part of the system, while the gradient of the maps as in Fig.~\ref{fig:dn-p1} steepens. However, we emphasize that the map within this setup is still continuous. In contrast, the derivative discontinuity refers to the discontinuous behavior of the gradient of the density maps along the cut of the particle number at integer value~\cite{Perdew1982}. The discontinuity is an exact concept for systems with degenerate ground state, where the maps are constructed as convex combination of the degenerate densities belonging to different particle number. The degeneracy of the eigenvalues of the ground state is due to an external potential within the Hamiltonian that serves as a Lagrange multiplier shifting the ground-state energy to states with different particle number. In the case of degeneracy, the derivative discontinuity shows up along the cut of the conjugated variable, e.g in purely electronic systems along $N$ or $\delta n$. 
We can conclude that the mapping becomes sharper for increasing electron-photon coupling strength $\lambda$ and therefore reminiscent to the case of static {electronic} correlation~\cite{Dimitrov2016}.
\begin{figure}[!htbp]
\centerline{\includegraphics[width=0.5\textwidth]{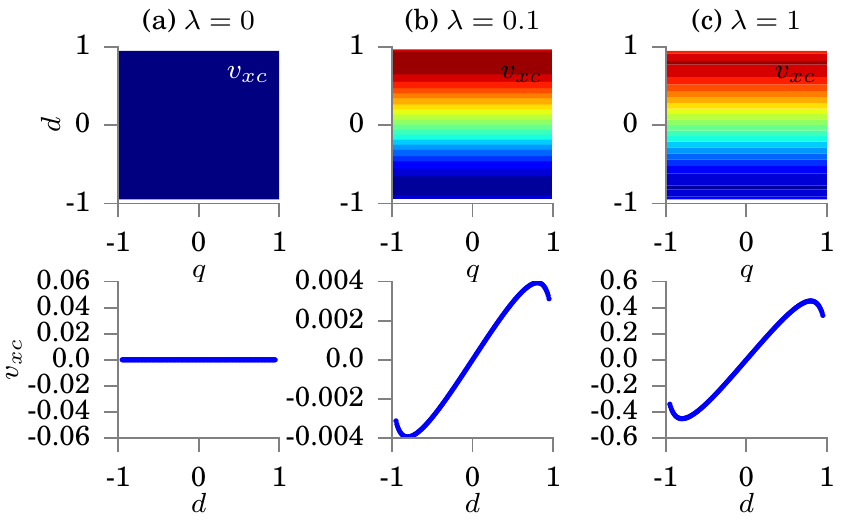}}
\caption{Single electron on two a two-site lattice: the xc potential $v_\text{xc}$ as function of the internal variables $(d,n)$ is shown in the first row. The second row shows the cut of ${v}_\text{xc}$ for $q=0$ for different coupling strength of (a) $\lambda=0$, (b) $\lambda=0.1$, and (c) $\lambda=1$.}
\label{fig:dv-p1}
\end{figure}
We plot the xc potential for this case in Fig.~\ref{fig:dv-p1}. In (a), we plot the two-dimensional plot for $\lambda=0$ and the the cut for $q=0$. Naturally, we find $v_{xc}=0$ for this case, since electrons and photons do not interact. The case for $\lambda=0.1$ is shown in (b). The cut along $q=0$ shown in the bottom reveals a smooth curve for $v_\text{xc}$ as function of $n$. If we compare to the density-to-potential map from Fig.~\ref{fig:dn-p1} (b), we find that $v_\text{xc}$ has the highest amplitude at the density values that show the highest derivative in the density-to-potential map. This is to be expected, since the non-interacting auxiliary system has a rather smooth behavior (see Fig.~\ref{fig:dn-p1} (a)), while the fully coupled problem is subject to the intra-system steepening, and consequently the xc potential functional has to compensate this mismatch. Thus the intra-system steepening directly translates to the size of the xc potential, which in the case of the two-site Rabi-Hubbard model implies a large potential step between the sites. This is a reminiscence of the step and peak structure of the photonic xc potential in full real space. In (c), we show the mapping for $\lambda=1$. For this case $v_\text{xc}$ has larger amplitudes in all regions, but its overall shape remains similar to the $\lambda=0.1$ case. We note, that such a scaling behavior could be employed to construct novel approximations to the xc potential. Further, we point out that the dependency of $v_\text{xc}$ on $q$ is below our numerical accuracy, thus very small in the considered parameter range. In general $q$ takes values from $-\infty$ to $\infty$ and in the case that $q$ takes such high values it will affect $v_{xc}$ more strongly. The $(d,q)$ behavior of the xc functional will be discussed in a little more detail at the end of this section. As a conclusion, we find that the steepening that is visible in Fig.~\ref{fig:dn-p1} along the new polaritonic coordinate $\tilde{v}_\text{ext}$ becomes here visible along $d$.\\
\begin{figure}[!htbp]
\centerline{\includegraphics[width=0.5\textwidth]{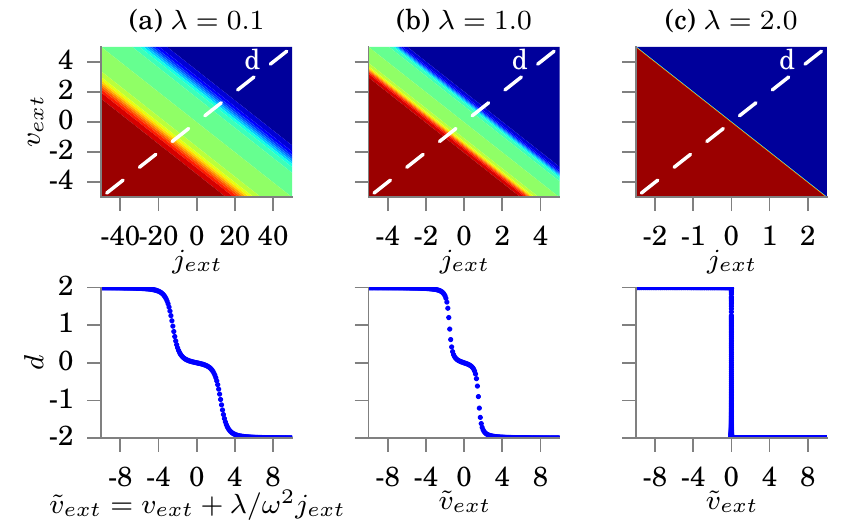}}
\caption{Two electrons with $U_0=5$ on a two-site lattice: the dipole $d$ as function of the external variables $(v_\text{ext},j_\text{ext})$ is shown in the first row. The second row shows the antidiagonal cut of $d(v_\text{ext},j_\text{ext})$ as indicated by the dashed line in the upper plot both for different coupling strength of (a) $\lambda=0.1$, (b) $\lambda=1$, and (c) $\lambda=2$.} 
\label{fig:dn-p2}
\end{figure}
Next, we analyze setup (ii), i.e., the two-site Rabi-Hubbard model in the two-electron subspace. The density-to-potential map is plotted in Fig.~\ref{fig:dn-p2}. In (a), we show the mapping for an electron-photon coupling strength of $\lambda=0.1$, hence a weak coupling setup. As in the case of the single electron, we also find here electron-photon correlation by the appearance of new normal modes. While the upper panel show the two-dimensional mapping ${d} [v_\text{ext},j_\text{ext}]$, in the lower panel, we show a antidiagonal cut along the new normal mode. The most noticeable difference to Fig.~\ref{fig:dn-p1} is that $d$ can now acquire values between $-2$ and $+2$ and in the mapping an intermediate step appears, where $d \approx 0$. This is, of course, due to the fact that we can now have two particles on one site and thus the total dipole moment can become $|2|$. If we now increase the electron-photon coupling strength $\lambda$ to $\lambda=1$, shown in Fig.~\ref{fig:dn-p2} (b), we find a steeper density-to-potential map. Also the intermediate step is reduced in size. In  Fig.~\ref{fig:dn-p2} (c), we plot the mapping for $\lambda=2$. Here, we find that the intermediate step vanishes and around $v_\text{ext}=j_\text{ext}=0$, the mapping becomes very steep. Since, we find approximately only two values for $d$, $-2$ and $+2$, meaning that both electrons are on the same side, we can conclude that the electron-photon interaction is capable of effectively reducing the electron-electron repulsion of the Hubbard term in Eq.~\ref{eq:ham_ph_mf}. Formulated differently, the electron-photon interaction mediates an effective attraction between the two electrons with the effect that both occupy the same site. Physically, we can interpret that the photons cloud the electrons such that the electron-electron repulsion is reduced. The static correlation of the electron-photon interaction dominates the correlation of the electron-electron interaction.
\begin{figure}[!htbp]
\centerline{\includegraphics[width=0.5\textwidth]{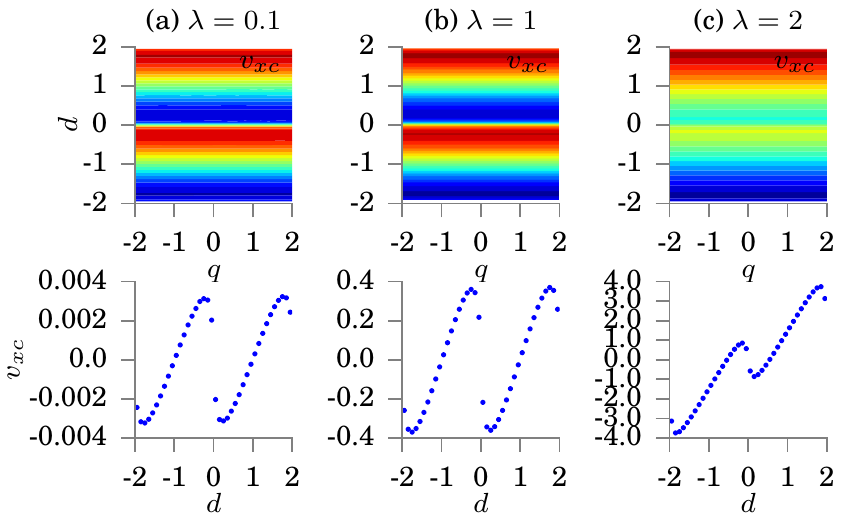}}
\caption{Two electrons with $U_0=5$ a two-site lattice: the xc potential $v_\text{xc}$ as function of the internal variables $(d,n)$ is shown in the first row. The second row shows the cut of ${v}_\text{xc}$ for $q=0$ for different coupling strength of (a) $\lambda=0.1$, (b) $\lambda=1$, and (c) $\lambda=2$.}
\label{fig:dv-p2}
\end{figure}
In Fig.~\ref{fig:dv-p2} we plot the $v_\text{xc}$ potential for the two electron case with different coupling strength. As in the case of a single electron, we find similar cuts for $v_\text{xc}$ for $q=0$ in (a) for $\lambda=0.1$ and in (b) for $\lambda=1$. Again, the intra-system steepening is responsible for the large values of the xc potential. In (c), where the coupling is increased to $\lambda = 2$, we find that due to the vanishing of the intermediate step, the regions of highest xc contributions are where the derivative due to the steepening is the largest, i.e., around $d=-2$ and $d=2$.
\\
\begin{figure}[!htbp]
\centerline{\includegraphics[width=0.5\textwidth]{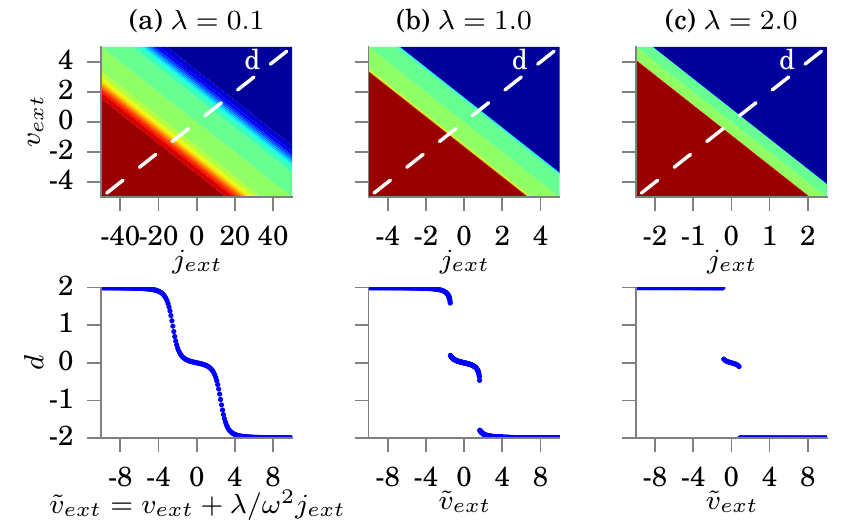}}
\caption{Two electrons with $U_0=5$ in mean-field approximation on a two-site lattice: the electron density ${d}$ as function of the external variables $(v_\text{ext},j_\text{ext})$ is shown in the first row. The second row shows the antidiagonal cut of ${d}(v_\text{ext},j_\text{ext})$ as indicated by the dashed line in the upper plot both for different coupling strength of (a) $\lambda=0.1$, (b) $\lambda=1$, and (c) $\lambda=2$.} 
\label{fig:dn-p2-mf}
\end{figure}
So far we have constructed the exact mappings. However, in practice we need to employ approximations since the exact mappings that constitute the Kohn-Sham potential are not known. Let us therefore see how the simplest approximate treatment of the coupled electron-photon problem, the afore introduced mean-field approximation of Eq.~\ref{eq:ham_e_mf2} performs. This will give us {insight about the missing xc potential}. In Fig.~\ref{fig:dn-p2-mf} (a), we plot the results in the regime of weak-coupling ($\lambda=0.1$). For the weak-coupling regime, we find a good agreement with the exact calculations shown in Fig.~\ref{fig:dn-p2}. The first differences become more pronounced in Fig.~\ref{fig:dn-p2-mf} (b). For the stronger coupling of $\lambda=1$, we find in comparison to Fig.~\ref{fig:dn-p2} (b) a broader intermediate step that is also less steep. The most significant differences are clearly visible in the strong-coupling limit for $\lambda=2$. While in Fig.~\ref{fig:dn-p2-mf} (c) we have seen the complete disappearance of the intermediate step, we find a remaining step if the classical approximation to the electron-photon coupling is employed. This clearly shows the breakdown of the classical approximation. Only in the limit of $\lambda\rightarrow \infty$, the classical approximation can correctly predict the vanishing  intermediate step. This brings us to the conclusion that this feature is a true electron-photon xc feature, where approximate xc functionals have to be developed to correctly account for such features. The missing electron-photon xc potential needs to enhance the steepening, i.e., it needs to model the missing correlation. This is in agreement with our interpretation of the intra-system steepening and correlation effects. The failure of the mean-field approximation in the strong-coupling limit around $\tilde{v}_\text{ext}\approx\pm2$ can be partially understood by comparing the exact eigenvalues versus the mean-field eigenvalues of our model system in the red-highlighted area in Fig.~\ref{fig:mf-eig}. For this setup, while the exact energy plotted in blue has a continuous and differentiable form, the mean-field energies develops a discontinuity in the red shaded area. How this discontinuity affects mean-field observables will be discussed in the next section.
In the remaining part of this section, we now study the implications of the features of the density-to-potential map on observables. As a consequence of the density map, in principle, arbitrary observables can be expressed in terms of the set of internal variables. In practice, however, the functional form of observables such as the photon number $N(q,d)$ is unknown and the functional development of important observables will push the framework of QEDFT to a practical level. While first functionals have been developed for simple model systems~\cite{flick2016a}, most functionals for observables remain unknown. For our model system, we can explicitly construct the dependency of selected observables on both, i.e. on the set of internal and external variables. Even though, the set of $(v_\text{ext},j_\text{ext})$ is mathematically equivalent to the set $(d,q)$,  the dependence on the set $(d,q)$ can be very different to the dependence on $(v_\text{ext},j_\text{ext})$.
The first observable we study is the interaction energy  $E_\text{int}$ that can be defined from Eq.~\ref{eq:ham_e_ph} by $E_\text{int} = -\omega \langle \hat{q} \, \hat{d}\rangle$. It is connected to the xc energy by 
\begin{align}
E_\text{xc} =E_\text{int} - E_\text{int, mf} = -\omega \left(\langle \hat{q} \, \hat{d}\rangle- q\, d \right)
\end{align}%
\begin{figure}[!htbp]%
\centerline{\includegraphics[width=0.5\textwidth]{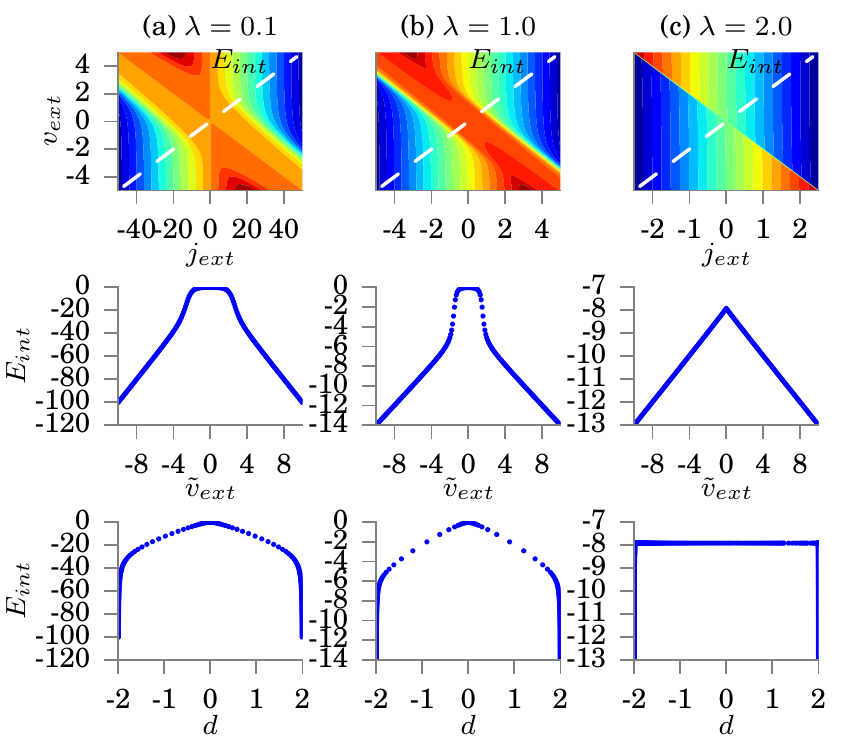}}%
\caption{Two electrons with $U_0=5$ on a two-site lattice: the interaction energy $E_{int} = \omega \langle \hat{q} \hat{d}\rangle$  as function of the external variables $(v_\text{ext},j_\text{ext})$ is shown in the first row. The second row shows the antidiagonal cut of $E_{int}(v_\text{ext},j_\text{ext})$ as indicated by the dashed line in the upper plot. The third row shows the diagonal cut of $E_{int}(v_\text{ext},j_\text{ext})$. All plots are shown for three different coupling strengths (a) $\lambda=0.1$, (b) $\lambda=1$, and (c) $\lambda=2$.}%
\label{fig:e-int-v-j}%
\end{figure}%
$E_{int}[v_\text{ext},j_\text{ext}]$ for the two-site Rabi-Hubbard model for two electrons is shown in Fig.~\ref{fig:e-int-v-j} and the corresponding observable in mean-field approximation is shown in Fig.~\ref{fig:e-int-v-j-mf}. In (a), the weak-coupling is shown, respectively. We find here the new normal coordinates and the intermediate step causes a distinguishable behavior around $j_{\text{ext}}\sim0$. This intermediate step becomes smaller for $\lambda=1$ shown in (b). In the strong-coupling limit, the interaction energy has a vanishing step in the exact solution of the problem shown in  Fig.~\ref{fig:e-int-v-j} (c). In contrast the mean-field solution fails to correctly reproduce the exact sharp feature of the interaction energy leading to large xc contributions. This failure can be explained by the discontinuity in the energy as discussed in Fig.~\ref{fig:mf-eig}.
The next observable, we study is the photon number in the system $\langle \hat{N} \rangle = \langle \hat{a}^\dagger\hat{a}\rangle$. In general, and in difference to electronic observables, such as ${d}$, the photonic observables are not restricted to integer values  due to its underlying bosonic nature in contrast to the fermionic number of particles. In Fig.~\ref{fig:N-ph} (a), we show $N$ as functional of the external potentials,  $N[v_\text{ext},j_\text{ext}]$. In (a), in the weak-coupling limit for $\lambda=0.1$, we find that the external potential $v_\text{ext}$ has no large overall influence on this observables and the harmonic nature of this observable is given by the external current $j_\text{ext}$. In the two lower panels, we plot the diagonal and the antidiagonal cut. Since the observable is unbound, we can excite very high photon numbers, up to $1200$ for the studied examples. Next in (b), we show the case for $\lambda=1.0$. Here, we find that the external potential $v_\text{ext}$ can alter this observable in cases, where $N$ is small. Around $j_\text{ext}\sim 0$, we find a funnel-type structure of this observable which is connected to the intermediate step of the density-to-potential mapping shown in Fig.~\ref{fig:dn-p2}. In (c), we show the strong-coupling limit for $\lambda=2$. Here, we find for the antidiagonal cut of $N[v_\text{ext},j_\text{ext}]$ map a sharp feature around $j_\text{ext}\sim 0$. Again this is connected to the sharp features in the density-to-potential map. Also the new normal mode is clearly visible along the antidiagonal.
 \begin{figure}[!htbp]
\centerline{\includegraphics[width=0.5\textwidth]{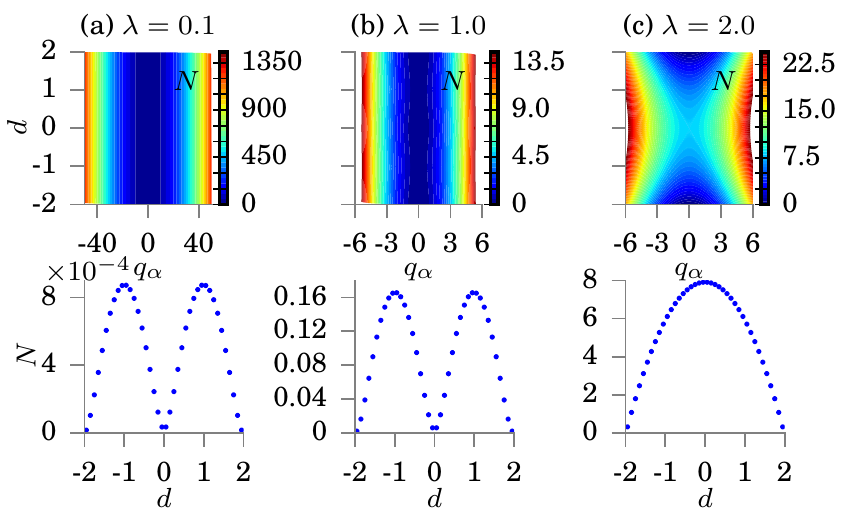}}
\caption{Two electrons with $U_0=5$ on a two-site lattice: the photon number $N=\langle\hat{a}^\dagger\hat{a} \rangle$ as function of the internal variables $({d},q)$ is shown for different coupling strength of (a) $\lambda=0.1$, (b) $\lambda=1$, and (c) $\lambda=2$ in the top and bottom for $q=0$.} 
\label{fig:N-ph-density}
\end{figure}
In Fig.~\ref{fig:N-ph-density}, we now show the dependency of  $N[{d}, q]$ on the internal variables in the top and in the bottom the cut for $q=0$. Here, we find that the appearing normal modes vanish for all three coupling strengths and the mapping becomes smooth. Qualitatively the weak-coupling $\lambda=0.1$ and the strong-coupling for $\lambda=1$ behave similarly (a double maximum in the cut), while the mapping for $\lambda=2$ has a constricted shape and only a single minimum in the cut. That the photon-number observable behaves more regularly when written in terms of the internal variables is an important detail. It suggests that we can find reasonable approximation to non-trivial functionals of the internal variables despite the intra-system steepening, which would make approximating much harder. Such non-trivial functionals are important to make QEDFT practical since in many situations it is not the density or the displacement field that one is interested in but rather, e.g., the energy or correlation functions of the photon field. We note that after changing to the internal variables, the dependency of $N[d,q]$ on $q$ becomes only strongly pronounced for high values of $q$. This implies that for a small amplitude of $q$, using functionals at $q=0$ becomes reasonable. This is very similar to the behavior we encountered in th xc potential functional. Also there the dependence of $v_\text{xc}$ on $q$ in the considered parameter range was very small. The weak dependence on only one parameter would not be the case if we used instead the mathematically equivalent external functional $v_\text{xc}[v_\text{ext}, j_\text{ext}]$ that would also allow to determine the dipole moment $d$ in the Kohn-Sham system. This is a nice example that the choice of the internal functional variables makes approximations much easier in practice.

\section{Four-site Rabi-Hubbard Model}

So far we have analyzed the simplest situation of electron-photon coupling and concluded that the intra-system steepening that appears in the density maps is a simple measure to quantify the electron-photon correlation. In this section, we now address the questions, whether the steepening also appears in more complex situations. To this end, we study a four-site Rabi-Hubbard model coupled to a single photon mode and demonstrate the implications of the discussed modifications of the density-to-potential map under strong light-matter coupling. We show how the density-potential map can help to find interesting behavior and explain experimentally observed effects. \\
The extension of Eq.~\ref{eq:ham_e_ph} to four sites is straightforward and the Hamiltonian for half-filling (four electrons) reads 
\begin{align}
   \hat{H}_0&=-t_0 \sum\limits_{i=1}^{3}\sum\limits_{\sigma=\uparrow,\downarrow}\left(\hat{c}_{i,\sigma}^\dagger\hat{c}_{i+1,\sigma} + \hat{c}_{i+1,\sigma}^\dagger\hat{c}_{i,\sigma}\right) \nonumber\\ &+U_0\sum\limits_{i=1}^4\hat{n}_{i,\uparrow}\hat{n}_{i,\downarrow}
   + \omega \hat{a}^{\dagger}\hat{a} - {\omega}\lambda \hat{q} \hat{d} +\frac{j_\text{ext}}{\omega}\hat{q}\nonumber\\
    &+ (\lambda\hat{d})^2/2 + v_\text{ext}\hat{d}
   \label{eq:ham_e_ph_4}
\end{align}
with $\hat{d} = d_0 \left( 3n_1 + n_2 - n_3 -3n_4  \right)$. In this case, $v_\text{ext}$ effectively is an external electric field, as routinely studied in electronic-structure calculations. For four sites, we construct the dipole to electric field map. Such a mapping of an reduced internal variable to an reduced external variable has been proven to be unique and has been analyzed e.g. in Ref.~\cite{Ruggenthaler2010}. Physically the gradient of the dipole moment to the external electric field describes the electric polarizability $\alpha$~\cite{Kuemmel2004}. In this spirit, we define the electric polarizability as follows
\begin{align}
\alpha[v_\text{ext}] = \frac{\delta d} {\delta \tilde{v}_\text{ext}},
\label{eq:polarizability}
\end{align}
where $\tilde{v}_\text{ext}$ describes the external electric field applied to the system as defined by Eq.~\ref{eq:ham_e_ph_4}. We note that for the two-site Rabi-Hubbard model studied in the previous section, the polarizability $\alpha$ is the gradient of the density-to-potential map. Thus, the larger the gradient in the mapping becomes, the larger values for the polarizability are obtained. In conducting polymers, it has been demonstrated that this high polarizability is directly connected to charge-transfer, i.e. conductivity~\cite{Mujica2000,Kuemmel2004,Mazinani2016}.\\
\begin{figure}[!htbp]
\centerline{\includegraphics[width=0.5\textwidth]{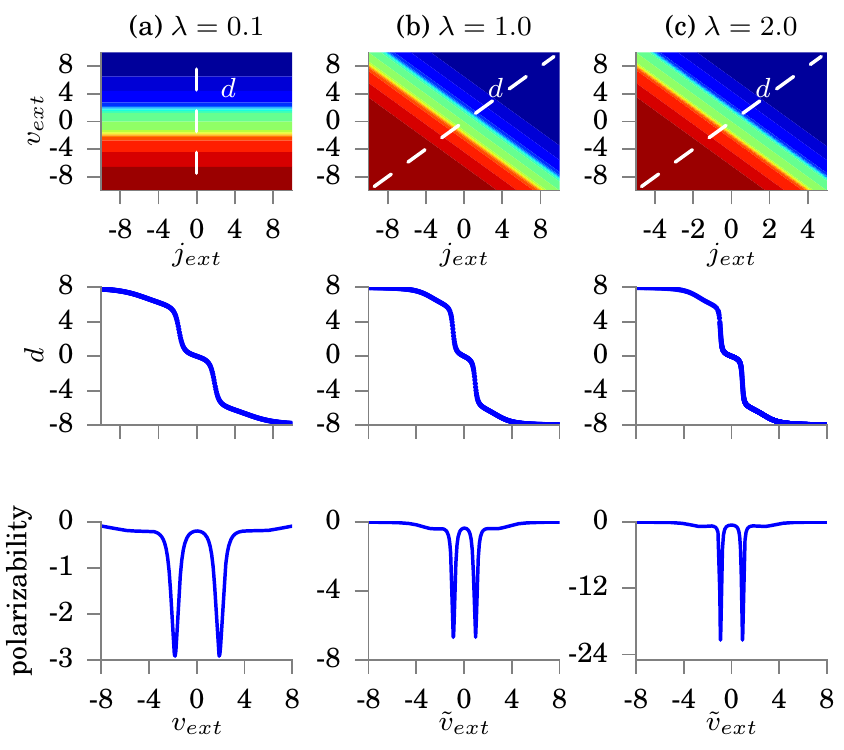}}
\caption{Four electrons with $U_0=2$ on a four-site lattice: the dipole moment $d$ as function of the electric field $\tilde{v}_\text{ext}$ is shown in the first row. The second row shows the polarizability $\alpha$ as defined in the main text. All plots are shown for three different coupling strengths (a) $\lambda=0$, (b) $\lambda=1$, and (c) $\lambda=2$.} 
\label{fig:polarizability-four-sites}
\end{figure}
In Fig.~\ref{fig:polarizability-four-sites}, we show how the electronic dipole moment $d$ and the polarizability $\alpha$ as function of the applied external potentials ${v}_\text{ext}$ and $j_\text{ext}$ change. Also in this more complex situation, we find new normal modes appearing. Thus, in Fig.~\ref{fig:polarizability-four-sites}, we show how $\tilde{v}_\text{ext}$ induces changes under strong light-matter coupling to the system. Without coupling, shown in (a), we find that the dipole moment develops three quasi-stationary regions, where the extremal values correspond to situations, where two electrons occupy the outermost sites and the other two electrons occupy the neighboring site. In the lower panel of Fig.~\ref{fig:polarizability-four-sites}, we plot the polarizability $\alpha$ as defined in Eq.~\ref{eq:polarizability}. We find two peaks in between the stationary regions of the dipole moment. If we now increase the electron-photon coupling, shown in (b) for the case of $\lambda=1$, we find that similarly as reported in the previous section, the dipole moment as function of the external potential steepens and the step around $v_\text{ext}\sim 0$ becomes narrower. Accordingly, the two peaks in the polarization shown in the bottom panel get close together and have larger amplitudes in comparison to the setup in (a). For strong-coupling that is here $\lambda=2$ shown in (c), we find that the middle step becomes even narrower and also the two peaks shown in the bottom panel become closer with high amplitude. In conclusion, we find that by tuning the electron-photon coupling strength, the polarizability of the system can be strongly influenced leading to a highly polarizable system.
\section{Summary and Outlook}
In this paper we have constructed the exact density-to-potential maps for electron-photon model systems and extended the concept of the intra-system steepening to general fermion-boson systems. We made explicit how the intra-system steepening can be used to identify large xc potentials and how these effects show up in other observables. We have identified the appearance of new normal modes in the coupled matter-photon system and showed how the density-to-potential maps can be constructed for all possible external pairs from only knowing the map along the polaritonic external potential $\tilde{v}_\text{ext}$. Finally we have highlighted for a four-site model with four electrons coupled to photons, how the intra-system steepening allows to identify interesting physical effects such as an increase of the polarizability of the matter system due to ultra-strong coupling to the photons. The increase in the polarizability is directly relevant for experiments such as in Ref.~\cite{orgiu2015}, where an increase in conductivity for organic semiconductors in strong coupling was measured.

The exact maps and the tools to analyze the importance of xc contributions will be helpful to further develop xc functionals for QEDFT that accurately capture the coupling between the charged particles and the photons. Also the finding that observables behave more regularly when represented by the internal variables is an important detail in the development of QEDFT. Such functionals become crucial for the practicability of QEDFT, as many observables are non-trivial functionals of the internal variables $n(\textbf{r})$ and $q_\alpha$, e.g., the number of photons. Their availability will allow for novel applications of density-functional methods in the context of quantum optics or plasmonics. Further, although the functionals in QEDFT are different to the ones of standard DFT, insights from a more complete description of real systems, i.e., also treating the photons, might prove beneficial also for DFT. Especially when going beyond the dipole approximation, the minimal-coupling prescription forces us to use the full current density to describe the coupling to the photon field. In this context a current-density functional (CDFT) scheme becomes unavoidable~\cite{Ruggenthaler2014, ruggenthaler2015b}. It seems possible by studying coupled matter-photon systems beyond the dipole approximation that we get novel insight also into CDFT. It would be very interesting to also investigate the exact density-to-potential maps for a Hubbard system that is coupled via its charge current to the photons, e.g., via a Peierls substitution. Such results would highlight the necessary ingredients of xc functionals to describe matter that only locally interacts strongly with photons, in contrast to the dipole approximation, where all electrons feel the same photon field. This would allow to calculate quantum local-field effects from first principles.
\section{Acknowledgements}
We thank Heiko Appel and Soren E. B. Nielsen for very fruitful discussions and acknowledge financial support from the European Research Council (ERC-2015-AdG-694097), and the European Union's H2020 program under GA no.676580 (NOMAD).
\bibliographystyle{apsrev4-1}
\bibliography{references}

\end{document}